# Band structure and decay channels of thorium-229 low-lying isomeric state for ensemble of thorium atoms adsorbed on calcium fluoride

**Petr V. Borisyuk[1], Oleg S. Vasilyev[1], Andrey V. Krasavin[1], Yury Yu. Lebedinskii[1,2], Victor I. Troyan[*,1], Eugene V. Tkalya[**,3,4]**

[1] National Research Nuclear University "MEPhI" (Moscow Engineering Physics Institute), Kashirskoye shosse 31, 115409 Moscow, Russia
[2] Moscow Institute of Physics and Technology (State University), Institutskiy per. 9, 141700 Dolgoprudny, Moscow Region, Russia
[3] Skobeltsyn Institute of Nuclear Physics, Lomonosov Moscow State University, Leninskie gory, Moscow 119991, Russia
[4] Nuclear Safety Institute of Russian Academy of Science, Bol'shaya Tulskaya 52, Moscow 115191, Russia

The results are presented on the study of the electronic structure of thorium atoms adsorbed by the liquid atomic layer deposition from aqueous solution of thorium nitrate on the surface of $CaF_2$. The chemical state of the atoms and the change of the band structure in the surface layers of $Th/CaF_2$ system on $CaF_2$ substrate were investigated by XPS and REELS techniques. It was found that REELS spectra for $Th/CaF_2$ system include peaks in the region of low energy losses (3-7 eV) which are missing in the similar spectra for pure $CaF_2$. It is concluded that the presence of the observed features in the REELS spectra is associated with the chemical state of thorium atoms and is caused by the presence of uncompensated chemical bonds at the $Th/CaF_2$ interface, and, therefore, by the presence of unbound 6d- and 7s-electrons of thorium atoms. Assuming the equivalence of the electronic configuration of thorium-229 and thorium-232 atoms, an estimate was made on the time decay of the excited state of thorium-229 nuclei through the channel of the electron conversion. It was found that the relaxation time is about 40 μs for 6d-electrons, and about 1 μs for 7s-electrons.

**1 Introduction** The low-lying nuclear isomeric level of $^{229}$Th isotope is not only a unique exception in nuclear physics with uncharacteristically low energy of $7.8 \pm 0.5$ eV [1], but it also is of exceptional interest due to many possibilities of use in various fields. The nuclear transition, shielded from the environment by the electron shell, is located in the vacuum ultraviolet region which is accessible by laser spectroscopy. This transition could serve as the basis of an optical frequency standard and could improve the accuracy of the existing standard on $Al^+$ ions by orders of magnitude [2]. The precise measurement of the energy of the isomeric transition will improve the accuracy of satellite navigation systems (GPS, GLONASS), directly related to the accuracy of frequency standards used; will make it possible to record the frequency dependence of the transition on the gravitational field, i.e., to measure the gravitational field of the Earth; will resolve a number of problems in fundamental physics, in particular, the measurement of the variation of some fundamental constants with high precision [3].

Another important use of the unique transition may be the development of gamma-laser of optical range [4], a fundamentally new device, in which the radiation is produced by medium with inverted population of atomic nuclei. Attempts to create such a device were taken during last fifty years [5, 6].

The value of the isomeric transition energy in [1] was obtained by indirect measurements; the direct registration of the transition is a major challenge, and still not realized [7]. This fact is caused by radioactivity of $^{229}$Th isotope, its absence in nature (it can be obtained with the use of nuclear reactions only), narrow spectral line and weak oscillator strength of the transition.

To solve the problem of measuring the energy of the isomeric transition, various mechanisms of excitation of the isomeric state were proposed, and a variety of physical systems containing $^{229}$Th isotope were used. One of the methods consists in placing $^{229}$Th ions into the ion trap [8]; mainly $^{229}$Th$^{+}$ and $^{229}$Th$^{3+}$ ions are used. $^{229}$Th$^{+}$ ions have a complicated system of electron levels, but are relatively easy to obtain and can be used for implementing the mechanism of electron bridge to excite the isomer state [9]; $^{229}$Th$^{3+}$ ions are the most convenient for finding the energies of the isomeric levels due to the high ionization potential (27 eV) and relatively simple system of electronic levels [3]. Another method consists in embedding $^{229}$Th ions into a crystal [10]; the quality factor of the nuclear transition which is sensitive to the additional influence caused by the presence of crystal structure, remains, nevertheless, much larger compared to any electronic transitions [11]. The advantage of using a crystal is the high density of $^{229}$Th nuclei, which is ~$10^{19}$ cm$^{-3}$ and is many orders of magnitude higher than the value achieved in the ion traps ($\leq 10^{8}$ nuclei [12]). This can greatly facilitate the direct observation of the transition. Suitable crystals should have transparency in the vacuum ultraviolet region. In addition, these crystals must admit $^{229}$Th$^{4+}$ ions in the exact locations of the crystal lattice to minimize the inhomogeneous broadening of the line. Such crystals are LiCaAlF$_6$, LiSrAlF$_6$, YLiF$_4$, CaF$_2$, Na$_2$ThF$_6$ [13, 14].

A crystal of CaF$_2$ doped with atoms of thorium-229 is the most simple in the view of experimental implementation. One of the techniques for obtaining such a crystal is a sequential use of atomic layer deposition and electron-beam deposition. The deposition of thorium atoms is carried out by chemical adsorption on the surface of CaF$_2$, and electron-beam deposition provides a CaF$_2$-film on top of the deposited layer of thorium atoms. After multiple repetition of this procedure a bulk CaF$_2$-crystal doped with thorium atoms can be obtained. However, such an implementation requires control of chemical and electronic properties of the surface at each step of deposition.

The paper presents the results of a study of the electronic structure of adsorbed thorium atoms formed when by atomic layer deposition from an aqueous solution of thorium nitrate on the surface of CaF$_2$. The chemical state of the atoms and the change of the band structure in the surface layers of Th/CaF$_2$ system were studied by X-ray photoelectron spectroscopy (XPS) and by reflection electron energy-loss spectroscopy (REELS). It was found that REELS spectra for Th/CaF$_2$ system include peaks in the region of low energy losses (3-7 eV) which are missing in the similar spectra for pure CaF$_2$. These peaks, apparently, are caused by the presence of the impurity levels of surface states in the forbidden band of CaF$_2$. It was established after heating of the sample and the detailed analysis of XPS and REELS spectra that the presence of these peaks is unambiguously related to the chemical state of thorium atoms adsorbed on the surface of the CaF$_2$ substrate, and is caused by the presence of uncompensated chemical bonds at the Th/CaF$_2$ interface, i.e. by the presence of unbounded 6d- and 7s-electrons. Assuming the equivalence of the electronic configuration of thorium-229 and thorium-232 atoms, an estimate was made on the time decay of the excited state of thorium-229 nuclei through the channel of the electron conversion. It was found that the relaxation time is about 40 µs for 6d-electrons, and about 1 µs for 7s-electrons.

Thus, the experimentally observed features of REELS and XPS spectra, combined with the estimates mentioned above show that for thorium-229 atoms adsorbed on the surface of CaF$_2$ the decay of the isomeric state will not be realized by the output of a gamma-quantum.

**2 Experimental technique** The formation of thorium atoms adsorbed on the surface of the calcium fluoride was performed by liquid atomic layer deposition at room temperature as a result of the surface reaction in aqueous solutions of precursors Th(NO$_3$)$_4$×5H$_2$O and HF. To this purpose a freshly cleaved surface of CaF2 crystal with dimensions of 6 mm×16 mm×1 mm (L×W×H) was maintained for 5 s in 0.1% aqueous solution of thorium nitrate, then the sample was moved into a vessel containing deionized water and was washed there for 5 s. After that the sample was placed in 2% aqueous solution of hydrofluoric acid and was kept there for 5 s for surface enrichment of missing links of fluorine. At the end of one cycle unbounded fluorine was removed by washing the sample in the vessel with deionized water for 5 s. According to this technique, the sample was prepared after 10 cycles with the surface concentration of thorium atoms ~ 0.1 ML (~ $10^{13}$ atoms of $^{232}$Th). The procedure of deposition is schematically presented in Fig. 1. Chemical vessels of chemical resistant fluoroplastic (PTFE) were manufactured specifically for this experiment and have not been used earlier. The reagents used in the experiments were chemically pure.

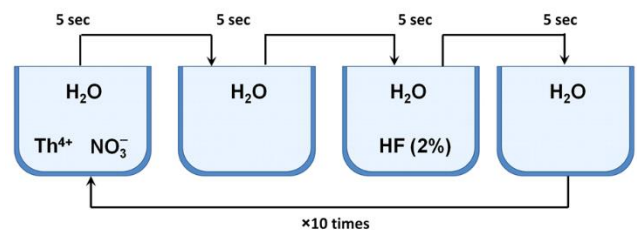

**Figure 1** Formation of system of thorium atoms adsorbed on the surface of calcium fluoride by liquid atomic layer deposition at room temperature as a result of the surface reaction in aqueous solutions of Th(NO$_3$)$_4$×5H$_2$O and HF precursors.

It should be noted that due to the high radiation activity of $^{229}$Th isotope the formation of monolayer thorium coatings was carried out with the salts of natural $^{232}$Th isotope only; that reduced the radioactivity of the sample to the permitted level under normal laboratory conditions.

The chemical composition and the electronic structure of thorium layers deposited on the surface of $CaF_2$ were monitored *ex situ* by XPS technique. For this purpose the sample was moved into ultrahigh vacuum (p ≈ 5×10$^{-9}$ Torr) chamber of energy analyzer of XSAM-800 surface system (Kratos, UK) immediately after the atomic layer deposition. X-rays of MgK$\alpha_{1,2}$ line with energy of $hv$ =1253.6 eV were used for excitation of photoelectrons. The experimental error in measuring the binding energy was 0.05 eV. The calibration of spectra was carried out according to the Ca2p line of the substrate with the binding energy of BE-$_{Ca2p3/2}$=347.8 eV [15].

The bandgap of the substrate and of layers formed on its surface was measured by REELS technique. The energy-loss spectra of scattered electrons were obtained *ex situ* by use of XSAM-800 spectrometer. The energy of the electron beam was $E_0$ = 500 eV; the beam current $I_0$ ≈ 30 μA; the scattering angle $\varphi_0$=125°±20° ($\Delta\varphi$=±20° is the angle of collecting of backscattered electrons). The energy spread of electrons of the primary beam after reflection was $\Delta E$≈ 1.5 eV.

Figure 2 shows the wide XPS and Auger spectra of the initial $CaF_2$-surface, and the surface after atomic layer deposition. The spectra obtained after deposition of thorium include Th (4f, 4d, 5d), Ca (2s, 2p, 3s, 3p), and F1s lines, and also C1s and O1s lines of low intensity, which indicates the presence of impurities (oxygen and carbon atoms) on the substrate surface.

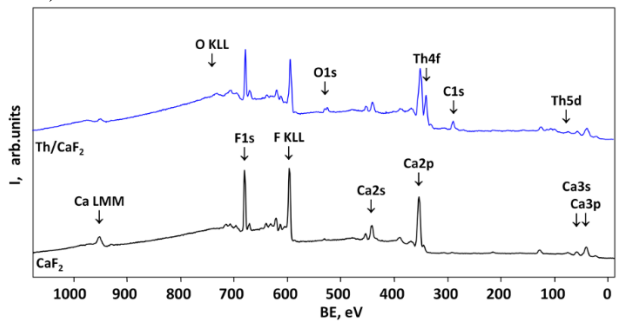

**Figure 2** The wide XPS and Auger spectra of the initial $CaF_2$-surface, and the surface after atomic layer deposition.

In order to analyze the electronic states and the stability of the chemical bonds of thorium atoms on the $CaF_2$-surface, the sample was annealed in vacuum. The heating was realized *in situ* in the chamber of the spectrometer analyzer. XPS and REELS measurements were conducted in the process of heating. The sample was kept for 10 minutes at each temperature before the measurement of the spectrum.

**3 Results** High resolution XPS spectra of Ca2p, Th4f and F1s lines obtained at different annealing temperatures are shown in Fig. 3. As the temperature increases, the intensity of the Ca2p$_{3/2}$ line from the substrate (the binding energy of 347.8 eV) remains unchanged for the calcium in a chemical bond with fluorine, while the intensity of Th4f line from adsorbed thorium atoms reduces, and the line broadens. The observed doublets Ca2p$_{3/2, 1/2}$ and Th4f$_{7/2, 5/2}$ are caused by the spin-orbit splitting. The position of the Th4f$_{7/2}$ line prior to the vacuum annealing (the temperature of the sample was 20°C) corresponds to the binding energy of 336.0 eV, which is close to the value for the bulk thorium tetrafluoride [15]. With the increase of temperature the Th4f$_{7/2}$ line shifts towards lower values of the binding energy up to 333.4 eV, which is close to the value of 333.1 eV for metal thorium [15].

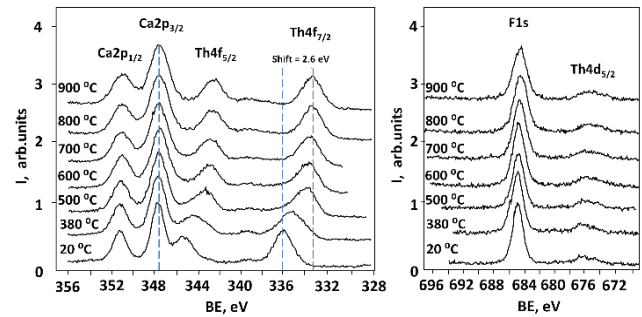

**Figure 3** XPS spectra of Ca2p, Th4f (a) and F1s (b) lines obtained of sub-monolayer thorium coatings at different annealing temperatures.

Together with the measurement of the XPS lines, REELS spectra were obtained in all stages of the annealing. Energy-loss spectra of electrons with energy of 500 eV scattered on the surface of pure $CaF_2$ and on the sample are shown in Fig. 4. The band gap for the bulk sample of calcium fluoride is 8.5 eV. There is an additional peak on REELS spectra of the test sample in the band gap region (energy loss range from 3 to 7 eV), which significantly changes during heating. This peak clearly indicates the emergence of large number of defects, which create local levels in the band gap, and can form a subzone at high concentration. On the other hand, the slight shift of the Th4f$_{7/2}$ line with respect to the Ca2p$_{3/2}$ observed in Fig. 3, indicates a change in the local potential caused by the chemical shift (spatial charge redistribution of valence electrons of thorium atoms caused by changes in the chemical bond with fluorine atoms leads to change of potential within the considered atom, and, consequently, to a change in the binding energy of the core levels [16]), and also a redistribution of contributions corresponding to different degrees of ionization of thorium atoms ($Th^{4+}$, $Th^{3+}$, $Th^{2+}$, $Th^{1+}$). Thus, the XPS Th4f$_{7/2}$ line with the binding energy of 336.0 eV may correspond to the superposition of $Th^{3+}$ and $Th^{4+}$ states, and the XPS Th4f$_{7/2}$ line with the binding energy of 333.4 eV may correspond to the superposition of $Th^{1+}$ and $Th^{2+}$ states.

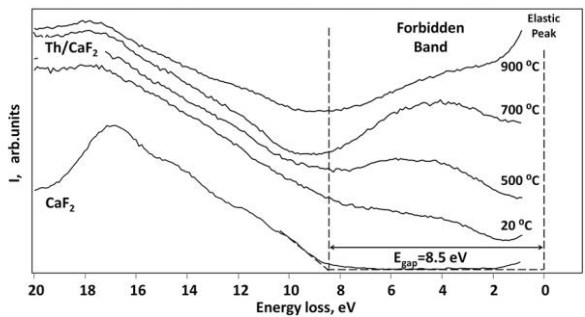

**Figure 4** Energy-loss spectra of electrons scattered on the surface of pure $CaF_2$ and on the sample at different annealing temperatures.

**4 Discussion** Based on the analysis of XPS and REELS spectra presented in Fig. 3 and Fig. 4, it can stated that the presence of peaks in the bandgap of calcium fluoride (even at room temperature) is unambiguously related to the chemical state of thorium atoms adsorbed on the surface of the $CaF_2$ substrate and, apparently, is caused by the presence of uncompensated chemical bonds at the $Th/CaF_2$ interface, resulting in the presence of unbounded 6d- and 7s-electrons of thorium atoms within the bandgap of the $CaF_2$ crystal.

However, the emergence of populated impurity levels formed within the bandgap of the $CaF_2$ crystal by 6d and 7s valence electrons of thorium may lead to appearance of additional decay channel of low-lying isomeric nuclear state $3/2^+$ ($E_{is}$=7.8±0.5 eV). If the energy gap $\Delta E$ between these states and the conduction band is less than the energy of the isomeric transition $E_{is}$, the process will take place of nonradiative transfer of excitation energy from the nucleus to an electron that leaves the filled subzone with kinetic energy of $E_e = E_{is} - \Delta E$.

There are two processes involving a nucleus and an electron in the second order of the perturbation theory for quantum electrodynamics which may be suitable for our case: an internal electron conversion at the isomeric transition in $^{229}$Th nucleus [17] and inelastic scattering of conduction electrons on excited $^{229}$Th nuclei [18]. Which of these processes is realized in this system depends on the properties of the impurity states. But for the states within the bandgap (i.e., when the impurity states lie several eV below the conduction band), the scattering process is impossible, since there are no free electrons in the subzone.

As for to the internal conversion, this process can actually take place, since the impurity levels in the $Th/CaF_2$ system retain a certain bond with thorium atoms and partly have quantum numbers of 6d- and 7s-states (amplitudes of 6d and 7s states have the weight close to 1 in the wave function of the impurity level).

In an isolated atom the internal electron conversion of nuclear M1 transition with energy of 7.8 eV has the probability of $W_{conv} \approx 10^6$ s$^{-1}$ for the 7s-shell of thorium, and $W_{conv} \approx 3 \times 10^4$ s$^{-1}$ for the 6d-shell. The calculation of the probability was made with use of code developed in [19] on the basis of the known code [20], and then advanced in [17]. The nuclear matrix element of the transition from the isomeric to the ground state $5/2^+$(0.0) was taken so that the reduced probability of the nuclear transition was $B_{W.u.}$(M1; $3/2^+$(7.8 eV)→$5/2^+$(0.0))=3×10$^{-2}$ in Weisskopf units [21].

Therefore, the life time of thorium atoms on the surface of the crystal with excited nuclei in the $3/2^+$(7.8 eV) state should be much less than 1 s. The high rate of decay is caused by the electronic conversion on electrons of filled impurity subzone.

It should be noted that the estimations made above are indirectly confirmed by the results of a recently published work [22]. In this study, a hard work has been done on the formation of $^{229}$Th/$CaF_2$ sample by chemical adsorption technique and excitation of isomeric nuclei. However, no signal was detected corresponding to the decay of the excited nuclei in the time range from tens of milliseconds to 1 s. This fact probably is caused by the conversion decay of $^{229}$Th$^m$ $3/2^+$(7.8 eV) nuclei. Excited thorium-229 nuclei transfer the excitation energy in nonradiative way to 6d- and 7s-electrons of the impurity subzone formed in the surface layer of $Th/CaF_2$ system. This explains the absence of nuclear decay photons.

A possible solution to this problem could be the formation of a bulk $CaF_2$ sample doped with thorium-229 nuclei. This would remove surface effects and compensate uncombined bonds of thorium atoms.

**Acknowledgements** This work was financially supported by Russian Foundation for Basic Research (project No. 14-08-00487a) and Ministry of Science and Education of Russia (project No. 3.1803.2014/K).